\documentclass[]{jfm}

\usepackage{amsmath,bm}

\usepackage{graphicx}
\usepackage{placeins}
\usepackage{subfig}
\usepackage[percent]{overpic} 
\usepackage{mathtools}
\usepackage{newtxtext}
\usepackage{newtxmath}
\usepackage{natbib}
\usepackage{hyperref}
\usepackage{marginnote}
\hypersetup{
    colorlinks = true,
    urlcolor   = blue,
    citecolor  = black,
}

\newcommand{\RomanNumeralCaps}[1]
\linenumbers

\usepackage{subfiles}

\newcommand{\bA}{\bm{A}}
\newcommand{\bB}{\bm{B}}
\newcommand{\bS}{\bm{S}}
\newcommand{\bu}{\bm{u}}
\newcommand{\dd}{{\rm d}}
\newcommand{\bg}{\bm {g}}

\newcommand{\ez}{\hat{\bm{e}}_z}

\title{Mechanism generating reverse buoyancy flux at the small scales of stably stratified turbulence}

\author{Soumak Bhattacharjee\aff{1},
  Stephen M. de Bruyn Kops\aff{2}
 \and Andrew D. Bragg \aff{1}
  \corresp{\email{andrew.bragg@duke.edu}}}

\affiliation{\aff{1} Department of Civil and Environmental Engineering, Duke University, Durham, NC 27708, USA
\aff{2} Department of Mechanical and Industrial Engineering, University of Massachusetts Amherst,
Amherst, MA 01003, USA}

\begin{document}
\maketitle

\begin{abstract}

Previous studies have shown that at the small-scales of stably stratified turbulence, the scale-dependent buoyancy flux reverses sign, such that there is a conversion of turbulent potential energy (TPE) back into turbulent kinetic energy (TKE) at these scales. Moreover, the magnitude of the reverse flux becomes stronger with increasing Prandtl number $Pr$. Using a filtering analysis we demonstrate analytically how this flux reversal is connected to the mechanism identified in Bragg \& de Bruyn Kops (JFM 2024 Vol 991 A10) that is responsible for the surprising observation that the TKE dissipation rate increases while the TPE dissipation rate decreases with increasing $Pr$ in stratified turbulence. The mechanism identified by Bragg \& de Bruyn Kops, which is connected to the formation of ramp-cliff structures in the density field, is shown to give the scale-local contribution to the buoyancy flux. At the smallest-scales this local contribution dominates and explains the flux reversal, while at larger scales a non-local contribution is important. Direct numerical simulations (DNS) of \textcolor{black}{3D} statistically stationary, stably stratified turbulence in the strongly stratified regime confirm the theoretical analysis, and indicate that while on average the local contribution only dominates the buoyancy flux at the smallest scales, it remains strongly correlated with the buoyancy flux at all scales. \textcolor{black}{The results show that ramp-cliffs are not only connected to the reversal of the local buoyancy flux but also the non-local part. At the small scales (approximately below the Ozmidov scale), ramp structures contribute exclusively to reverse buoyancy flux events, whereas cliff structures contribute to both forward and reverse buoyancy flux events.}

\end{abstract}


{\bf MSC Codes }  {\it(Optional)} Please enter your MSC Codes here

\section{Introduction}\label{sec:Introduction}

The study of turbulence in stably stratified fluids is vital for understanding and modeling atmospheric and oceanic flows.
The background mean density gradient can arise due to temperature and/or salinity gradients throughout the fluid, and leads to buoyancy forces that inhibit vertical motion and generate spatially intermittent and layered flow structures that are of tremendous consequence, yet difficult to understand \citep{Harindra1991,Staquet1996,RileyLindborg_2012,Caulfield2025}.

The relative diffusivity of the scalar on which density depends is quantified by the Prandtl number $Pr\equiv \nu/\kappa$, where $\nu$ and $\kappa$ are the momentum and the scalar diffusivities, respectively. 
For thermally stratified air $Pr \approx 0.7$, for thermally stratified water $Pr \approx 7$, whereas $Pr\approx 700$ for salt stratified water. Despite this wide range, much of the literature on stratified turbulence has focused on $Pr=1$, as also noted by \cite{Okino2020}, primarily because direct numerical simulations (DNS) with $Pr>1$ are $Pr^{3/2}$ times more computationally expensive (in terms of spatial resolution requirements) compared to $Pr \approx 1$ flows.

\cite{Gerz1989} observed that stratified, turbulent shear flows feature a counter-gradient (reverse) heat flux when the stratification is strong for $Pr=5$ but not for $Pr=0.7$.
\cite{Komori1996} compared grid generated turbulence using thermally stratified water at $Pr = 6$ with that using salt-stratified water at $Pr = 600$ and found that higher-Prandtl-number fluids develop stronger, smaller-scale reverse buoyancy fluxes. In their recent study of forced, stratified turbulence using DNS with $0.7 \leq Pr \leq 8$ and for varying stability strengths, \cite{Legaspi2020} observed that at high-wavenumbers the buoyancy flux becomes positive, indicating conversion of potential energy to kinetic energy at small-scales, which had also been discussed in \cite{Holloway1988,BA_1996,Carnevale2001}.
While different arguments have been given for the origin of the reverse  buoyancy flux, there is no general consensus on the mechanism that leads to its development or $Pr$ dependence.  
For example, \cite{Holloway1988} suggests that non-linear interactions transfer potential energy downscale more effectively than it does kinetic energy; this leads to a relative excess of potential energy which is converted to kinetic energy through the buoyancy flux, also discussed in \cite{Legaspi2020}. 
The plausibility of this argument is, however, unclear. 
Indeed, if kinetic energy is transferred downscale less effectively than potentially energy, this would seem to suggest that kinetic energy might pile up at some scales, not potential energy, and some of this could then be transferred to the potential field. But this is the opposite of what was suggested in \cite{Holloway1988}.

\cite{RileyCdBKops_2023} recently investigated the decay of a stably-stratified turbulence and observed that higher $Pr$ leads to a smaller mixing coefficient $\Gamma\equiv \langle\chi\rangle/\langle\epsilon\rangle$, where $\langle\chi\rangle$ and $\langle\epsilon\rangle$ are the average turbulent potential and kinetic energy dissipation rates, respectively \citep[c.f.][]{Okino2019}. 
In order to explain this, \cite{BraggdBKops_2024} analysed the equations governing the velocity and density gradients in stratified turbulence, which describe $\langle\chi\rangle$ and $\langle\epsilon\rangle$, and they identified a new mechanism that explains why $\Gamma$ decreases with increasing $Pr$. According to this mechanism, the emergence of ramp-cliff structures in the density field causes the mean density gradient to oppose the production of fluctuating density gradients (reducing $\langle\chi\rangle$), and support the growth of velocity gradients (increasing $\langle\epsilon\rangle$). They also showed, using asymptotic analysis, that this effect should grow with increasing $Pr$ (although the effect will saturate at sufficiently large $Pr$), and this is why $\Gamma$ reduces as $Pr$ increases. Results from DNS (using the same database as \cite{RileyCdBKops_2023}) confirm these arguments. 
\textcolor{black}{The objective of this paper is to establish the connection between the gradient field dynamics, along the lines noted by \cite{BraggdBKops_2024}, with the small-scale buoyancy flux reversal that has been observed in previous studies of stratified turbulence. 
Should such a connection exist, it would provide a mechanistic explanation for the occurrence of a small-scale buoyancy flux reversal, linking it to ramp–cliff structures, a fundamental feature of of scalar turbulence driven by a mean scalar gradient.}

The rest of the paper is organised as follows: in \S 2 we present our theoretical analysis that establishes the connection,
in \S  3 we summarise the DNS used, and in \S  4 we present and discuss the
results from our DNS and consider them in view of the theoretical analysis. Finally, in \S  5 we summarise the findings and discuss future work.

\section{Theory}\label{sec:Theory}



We consider a stably stratified flow with fluctuating $\phi$ and mean $\Phi_b=(\rho_r + \gamma z)g/(N \rho_r)$ densities (that have been normalized so that they have the dimensions of a velocity), where $\rho_r$ is the reference density, $\gamma<0$ is the constant, mean density-gradient, $g$ is the magnitude of the gravitational acceleration, and $N\equiv\sqrt{-g \gamma/\rho_r}$ is the Br\"unt-V\"ais\"al\"a frequency.
Note that the mean density gradient $\gamma \ez$ points in the direction of gravity $\hat{\bm{g}} = -\ez$ since $\gamma<0$.
The Navier-Stokes-Boussinesq equations are given by
\begin{align}
D_t \bu &=-\left(1 / \rho_r\right) \nabla p+ 2 \nu \nabla \bm{\cdot} \bm{S}-N \phi \ez+\bm{F}, \label{eq:NSE_boussinesq}\\
D_t \phi &= \kappa \nabla^2 \phi+N u_z,\label{eq:phi}
\end{align}
where $D_t \equiv \partial_t + {\bu} \,\bm{\cdot}\bm\nabla \,$ is the material derivative operator, $\bu$ is the velocity, $\nu$ is the kinematic viscosity, $\kappa= \nu/Pr$ is the scalar diffusivity,
$\bm{S}\equiv(\bA + \bA^\top)/2$ is the strain-rate field, and $\bA\equiv\nabla \bu$ is the velocity-gradient field. $\bm{F}$ is a forcing term which is used in the DNS to generate an approximately statistically stationary state (details on this are given later).

The flow can be characterised using the Reynolds number $Re \equiv UL/\nu$ and the Froude number $Fr \equiv U/(LN)$, where $U$ and $L$ are the horizontal r.m.s. velocity and horizontal integral length scale of the flow.
The smallest spatial scale (in a mean-field sense) of the velocity field is the Kolmogorov scale $\eta \equiv (\langle \epsilon \rangle^3/\nu)^{1/4}$, while the smallest scale in the density field is the Bachelor scale, which is related to $\eta$ through $\eta_B = \eta/Pr^{1/2}$ when $Pr\geq 1$ \citep{Batchelor1959}. 

We employ a filtering approach to explore the dynamics of $\bm u$ and $\phi$ at different scales. The filter operator acting on an arbitrary field $a(\bm{x},t)$ is defined as
\begin{equation}\label{eq:def_filtering}
    \tilde{a}^\ell(\bm{x},t) \equiv \int a(\bm{y},t)\mathcal{G}_{\ell}(\bm{x}-\bm{y})\,d\bm{y},
\end{equation}
where $\tilde{a}^\ell$ contains the field information at scales $\geq \ell$, whereas the sub-\textcolor{black}{filter} scale information is captured by $a(\bm{x},t)-\tilde{a}^\ell(\bm{x},t)$. For notational simplicity, the $\ell$ in the superscript will be dropped and $\tilde{a}$ will denote a field filtered at a scale $\ell$, unless indicated otherwise.


\textcolor{black}{Following \cite{Germano_1992}, one can partition the energy associated with the velocity and density fluctuations into filter scale energy and sub-filter scale energy fields, the latter being the focus here.
The sub-filter turbulent kinetic energy (TKE) is given by $e_{K}\equiv(\widetilde{\|\bu\|^2}-\tilde{\bu} \bm{\cdot} \tilde{\bu})/2$, whereas the sub-filter turbulent potential energy (TPE) is given by 
$e_P\equiv (\widetilde{\phi\phi}-\tilde{\phi}\tilde{\phi})/2$.
The averaged equations for $e_{K}$ and $e_{P}$ in homogeneous, stably stratified turbulence are given by
\begin{align}
    \partial_t \langle e_{K} \rangle &= \langle \Pi_{K}\rangle + \langle \mathcal{B}\rangle - \langle \varepsilon_{K}\rangle+\langle\mathcal{F}_{K}\rangle \label{eq:SGStke_budget_ss1}\\
    \partial_t \langle e_\phi \rangle &= -\langle \mathcal{B}\rangle +\langle \Pi_\phi\rangle -\langle\varepsilon_\phi\rangle \label{eq:SGStke_budget_ss3}.
\end{align}
Here, $\Pi_{K}\equiv-(\widetilde{\bu\bu}-\tilde{\bu}\tilde{\bu}) \bm{:} \tilde{\bS}$ 
and $\Pi_\phi\equiv-(\widetilde{\bu \phi}-\tilde{\bu} \tilde{\phi}) \bm{\cdot} \tilde{\bB}$ are the scale-to-scale TKE and TPE fluxes ($\bB=\nabla \phi$ being the density gradient), $\varepsilon_{K}~\equiv~\nu \left(\widetilde{\|\nabla \bu\|^2} - \|\nabla \tilde{\bu}\|^2\right)$ and $\varepsilon_\phi \equiv (\nu/Pr) \left(\widetilde{\|\bB\|^2} - \|\tilde{\bB}\|^2\right)$ are the sub-filter TKE and TPE dissipation rates, and}
\begin{align}\label{eq:buoyancy_flux}
    \mathcal{B}\equiv-N \left( \widetilde{u_z \phi} - \tilde{u}_z \tilde{\phi}\right),
\end{align}
\textcolor{black}{is the sub-filter buoyancy flux} which couples the \textcolor{black}{sub-filter} TKE and TPE equations, and is defined such that $\mathcal{B}<0$ corresponds to a transfer from the sub-\textcolor{black}{filter} TKE field to the sub-\textcolor{black}{filter} TPE field.

For a statistically stationary flow, at scale  $\ell \gg L$, we have the balance $-\langle \mathcal{B} \rangle \sim \langle \chi \rangle>0$. At scales larger than the Ozmidov scale $\ell > l_{O}\equiv (\langle \epsilon \rangle/N^3)^{1/2}$ where buoyancy plays a leading order role in the flow energetics, one expects $\langle\mathcal{B}\rangle<0$ corresponding on average to a transfer of TKE into TPE. Based on the standard theory of stratified turbulence, the buoyancy \textcolor{black}{flux} is supposed to be negligible compared with the energy flux terms at scales $\ell < l_{O}$, and Kolmogorov-like turbulence is expected to emerge \citep{RileyLindborg_2012}. While this seems to be true for $Pr=O(1)$, recent arguments suggest it may not apply when $Pr\gg 1$ \citep{Bhattacharjeeetal2025}.

We now want to establish the connection between $\mathcal{B}$ and the gradient field dynamics analysed by \cite{BraggdBKops_2024}. 
From Eq. \ref{eq:NSE_boussinesq}, \ref{eq:phi} we can obtain equations governing  \textcolor{black}{the evolution of the averaged velocity and density gradient magnitudes, $\|{\bA}\|^2$ and  $\|{\bB}\|^2$, which (using homogeneity of the flow) are given by
\begin{align}
 \partial_t \langle \| {\bA}\|^2 \rangle &=  \langle \mathcal{P}_{A1}\rangle  - \langle \mathcal{P}_{B2}\rangle  - \langle \mathcal{D}_A \rangle  + \langle {\bA} \bm{:} \nabla \bm{F} \rangle, \label{eq:eps_budget}\\
\partial_t\langle \| {\bB}\|^2 \rangle &= \langle \mathcal{P}_{B1}\rangle  + \langle \mathcal{P}_{B2}\rangle  - \langle \mathcal{D}_B \rangle \label{eq:chi_budget}. 
\end{align}
}
In these equations\textcolor{black}{, $\bA$ and $\bB$ are the fluctuating velocity and density gradient fields, and the fluctuating productions terms are given by
\begin{align}\label{eq:PA1l_PB1l}
    \mathcal{P}_{A1} \equiv - \bA^\top \bm{:} {\bA} \bm{\cdot} {\bA}, \; \mathcal{P}_{B1} \equiv - \bB \bm{\cdot} {\bA}^\top \bm{\cdot} {\bB},
\end{align}
whereas the mean} production term arising from the \textcolor{black}{background} density gradient is
\begin{align}\label{eq:PB2}
    \mathcal{P}_{B2} \equiv N {\bB} \bm{\cdot} {\bA}^\top \bm{\cdot} \ez,
\end{align}
and $\mathcal{D}_A$, $\mathcal{D}_B$ are dissipation rates of the velocity and scalar gradient magnitudes \citep[see below Eq. 2.13, 3.1 for expressions]{BraggdBKops_2024}.
\cite{BraggdBKops_2024} noted that ramp-cliff structures in the field $\phi$ imply that $\bB$ exhibits the largest fluctuations when it points in the direction of the mean density gradient $\hat{\bg}$. 
\textcolor{black}{Further, as a consequence of $\langle \bB\rangle=\bm 0$ (since $\bB$ is the gradient of the fluctuating density), $\bB$ is preferentially misaligned with the mean density-gradient direction $\hat{\bg}$ and so
\begin{equation*}
    \mathbb{P}(\bB \bm{\cdot} \hat{\bg} >0) < \mathbb{P}(\bB \bm{\cdot} \hat{\bg} <0) \text{, or equivalently, }\mathbb{P}(B_z >0) > \mathbb{P}(B_z <0).
\end{equation*}
where $\mathbb{P}(\cdot)$ denotes the probability.} From this, they argued that $\langle \mathcal{P}_{B1} \rangle$ and $\langle \mathcal{P}_{B2} \rangle$ should have opposite signs\textcolor{black}{.
Moreover}, since $\langle \mathcal{P}_{B1} \rangle+\langle \mathcal{P}_{B2} \rangle>0$ at steady state (to balance the dissipation term \textcolor{black}{$\langle \mathcal{D}_B \rangle$ in \eqref{eq:chi_budget}}), it follows that $\langle \mathcal{P}_{B2} \rangle <0$ provided $N^2/\langle\| {\bB}\|^2\rangle<O(1)$ (i.e. the fluctuating density gradients are larger than the mean density gradient, which will be satisfied if the flow is turbulent unless $Pr$ is very small). 
The negativity of $\langle \mathcal{P}_{B2} \rangle$ means that it acts as a source for $\langle \epsilon \rangle = \langle \| {\bA}\|^2\rangle/\nu$ \eqref{eq:eps_budget} but as a sink for $\langle \chi \rangle=\langle \| {\bB}\|^2 \rangle Pr/\nu$ \eqref{eq:chi_budget}. 

\textcolor{black}{Following similar steps, one can derive the evolution of the filtered velocity and density gradient magnitudes,}
$\| \tilde{\bA}\|^2$ and  $\| \tilde{\bB}\|^2$\textcolor{black}{, for a homogeneous flow:
\begin{align}
 \partial_t \langle \| \tilde{\bA}\|^2 \rangle &=  \langle \mathcal{P}^\ell_{A1}\rangle  - \langle \mathcal{P}^\ell_{B2}\rangle  - \langle \mathcal{D}^\ell_A \rangle  + \langle \tilde{\bA} \bm{:} \nabla \bm{F} \rangle - \langle \tilde{\bA} \bm{:} \nabla \nabla \bm{\cdot} \bm{\tau}\rangle, \label{eq:filteredeps_budget}\\
\partial_t\langle \| \tilde{\bB}\|^2 \rangle &= \langle \mathcal{P}^\ell_{B1}\rangle  + \langle \mathcal{P}^\ell_{B2}\rangle  - \langle \mathcal{D}^\ell_B \rangle - \langle \tilde{\bB} \bm{\cdot} \nabla \nabla \bm{\cdot} \bm{\Sigma}\rangle \label{eq:filteredchi_budget}, 
\end{align}
where, the productions terms, $\mathcal{P}^\ell_{A1}$, $\mathcal{P}^\ell_{B1}$ and $\mathcal{P}^\ell_{B2}$
\begin{align}\label{eq:PA1B1B2_l}
    \mathcal{P}^\ell_{A1} \equiv - \tilde{\bA}^\top \bm{:} \tilde{\bA} \bm{\cdot} \tilde{\bA}, \; \mathcal{P}^\ell_{B1} \equiv - \tilde{\bB} \bm{\cdot} \tilde{\bA}^\top \bm{\cdot} \tilde{\bB}, \; \mathcal{P}^\ell_{B2} \equiv N \tilde{\bB} \bm{\cdot} \tilde{\bA}^\top \bm{\cdot} \ez
\end{align}
correspond to the filtered version production terms $\mathcal{P}_{A1}$, $\mathcal{P}_{B1}$ and $\mathcal{P}_{B2}$  (Eq. \ref{eq:eps_budget}, \ref{eq:chi_budget}).
In the limit $\ell/\eta_B \to 0$, Eq. \ref{eq:filteredeps_budget},  \ref{eq:filteredchi_budget} reduces to \eqref{eq:eps_budget} and \eqref{eq:chi_budget} respectively,  the filtered productions terms tend to the corresponding unfiltered values and the scale-to-scale flux terms (last terms in Eq. \ref{eq:filteredeps_budget},  \ref{eq:filteredchi_budget}) go to 0:
\begin{align*}
  & \tilde{\bA} \to \bA,\;\tilde{\bB} \to \bB,\quad  \mathcal{P}^\ell_{A1} \to \mathcal{P}_{A1},\; \mathcal{P}^\ell_{B1} \to \mathcal{P}_{B1},\; \mathcal{P}^\ell_{B2} \to \mathcal{P}_{B2},\\
& \tilde{\bB} \bm{\cdot} \nabla \nabla \bm{\cdot} \bm{\Sigma} \to 0,\; \tilde{\bA} \bm{:} \nabla \nabla \bm{\cdot} \bm{\tau} \to 0.
\end{align*}
}
Since the term $\mathcal{P}^\ell_{B2}$ comes from taking the gradient of the buoyancy term in Eq. \ref{eq:NSE_boussinesq}, there must be a connection between it and $\mathcal{B}$. 

\textcolor{black}{\citet{Johnson_2020,Johnson_2021} showed that for the special case of an isotropic, Gaussian filter kernel $\mathcal{G}_\ell$, an exact solution for the sub-filter stress $\widetilde{\bu\bu}-\tilde{\bu}\tilde{\bu}$ (that appears in the filtered TKE equation) in terms of the velocity gradients can be obtained as the solution to a forced diffusion equation whose Green's function is $\mathcal{G}_\ell$.} \textcolor{black}{The same analytical procedure can also be used to obtain an exact solution for the buoyancy flux $\mathcal{B}$ in terms of velocity and density gradients, yielding}
\begin{align}
\mathcal{B} &=  \textcolor{black}{\mathcal{B}^{{\rm l}}} +  \textcolor{black}{\mathcal{B}^{{\rm nl}}} \label{eq:partition_PB},\\
\textcolor{black}{\mathcal{B}^{{\rm l}}} &\equiv \textcolor{black}{- \ell^2 \mathcal{P}^\ell_{B2}} \label{eq:partition_Bl},\\
 \textcolor{black}{\mathcal{B}^{{\rm nl}}} &\equiv  -N \int_0^{\ell^2}  \boldsymbol{T}^\theta \left( \tilde{\bB}^{\sqrt{\alpha}}, \widetilde{\bA^\top}^{\sqrt{\alpha}} \right) \bm{\cdot} \hat{\bm{e}}_z\,\dd \alpha \label{eq:partition_Bnl},
\end{align}
where $\theta \equiv \sqrt{\ell^2 - \alpha}$ and $\boldsymbol{T}^\theta(\bm{a},\bm{b})\equiv\widetilde{\bm{a\cdot b}}^\theta - \tilde{\bm{a}}^\theta \bm{\cdot} \tilde{\bm{b}}^\theta$, for arbitrary tensors $\bm{a},\bm{b}$, and $ \textcolor{black}{\mathcal{B}^{{\rm nl}}} $ is the contribution \textcolor{black}{to the sub-filter-scale buoyancy flux from scales $< \ell$} (and we have shown explicitly the superscript denoting filtering at scale\textcolor{black}{s} $\theta$ \textcolor{black}{, $\sqrt{\alpha}$} in keeping with the definition from \eqref{eq:def_filtering}). 
In keeping with the terminology used in \citet{Johnson_2020,Johnson_2021}, the term \textcolor{black}{$\mathcal{B}^{{\rm l}}\equiv$} represents the ``scale-local'' contribution to $\mathcal{B}$ which \textcolor{black}{only involves contributions} from scales $\textcolor{black}{\geq}\ell$, while $ \textcolor{black}{\mathcal{B}^{{\rm nl}}} $ is the ``non-local'' contribution \textcolor{black}{which only involves contributions from} scales $< \ell$ (the sub-\textcolor{black}{filter} scales) on the energetics of the filtered fields.
%
The filtered term $\langle\mathcal{P}^\ell_{B2}\rangle$ was also analysed in \cite{BraggdBKops_2024} and it was argued that while it is negative at small scales since $\lim_{\ell/\eta_B\to 0}\langle\mathcal{P}^\ell_{B2}\rangle\to \langle\mathcal{P}_{B2}\rangle$, and $\langle\mathcal{P}_{B2}\rangle<0$, it must become positive at large-scales for a statistically stationary flow because this term must balance the sub-\textcolor{black}{filter} flux (last term in Eq. \ref{eq:filteredchi_budget}) which is expected to be positive (corresponding to a downscale flux of TPE). 
The result in \eqref{eq:partition_PB} then suggests that the mechanism discovered in \cite{BraggdBKops_2024} that generates $\langle\mathcal{P}_{B2}\rangle<0$ may also explain the change in the sign of the buoyancy flux $\langle\mathcal{B}\rangle$ at small-scales that has been observed in numerous previous studies. 
This is certainly the case for $\ell\leq O(\eta_B)$ where $\textcolor{black}{\mathcal{B}^{{\rm nl}}} $ will be sub-leading compared with $- \ell^2 \mathcal{P}^\ell_{B2}$ in \eqref{eq:partition_PB} because $\lim_{\ell/\eta_B \to 0} \textcolor{black}{\mathcal{B}^{{\rm nl}}} \to 0$. 
In this case, we may say that just as the formation of ramp-cliff structures is responsible for $\langle\mathcal{P}_{B2}\rangle$ becoming negative, they are also responsible for the emergence of an inverse buoyancy flux $\langle\mathcal{B}\rangle>0$ at scales $\ell\leq O(\eta_B)$. 
Whether the behaviour of $\langle\mathcal{P}^\ell_{B2}\rangle$ at scales $\ell> O(\eta_B)$ can also explain the behaviour of $\langle\mathcal{B}\rangle$ at these scales (and in particular its sign) will depend on how large the contribution from $\langle \textcolor{black}{\mathcal{B}^{{\rm nl}}} \rangle$ is to $\langle\mathcal{B}\rangle$. 

Due to the complexity of the integral defining $ \textcolor{black}{\mathcal{B}^{{\rm nl}}} $, estimates for its size relative to $\mathcal{B}$ are challenging. However, its behaviour in the limit $\ell/L\to \infty$ is known. In particular, for the statistically homogeneous system we are considering, $\tilde{\bA}$, $\tilde{\bB}$ and therefore $\mathcal{P}_{B2}$ all vanish in the limit $\ell/L\to \infty$ because in this limit $\tilde{\bA}=\overline{\bA}=\bm{0}$ and $\tilde{\bB}=\overline{\bB}=\bm{0}$, where $\overline{\cdot}$ denotes a spatial average over the domain of the flow. However, in this same limit, $\mathcal{B}=-N\overline{u_z\phi}\neq 0$ and hence in this limit we must have $\mathcal{B}^{{\rm nl}}=\mathcal{B}$. We will explore later the contribution from $\mathcal{B}^{{\rm nl}}$ across scales using DNS, \textcolor{black}{and the role that ramp-cliffs play in determining the sign of this term.}

Another important conclusion that follows from the behaviour of \eqref{eq:partition_PB} concerns the $Pr$-dependence of $\langle\mathcal{B}\rangle$. In particular, \cite{BraggdBKops_2024} showed that the magnitude of $\langle\mathcal{P}_{B2}\rangle$ grows with increasing $Pr$ (although it saturates at sufficiently large $Pr$), and therefore according to \eqref{eq:partition_PB}, so also must $\langle\mathcal{B}\rangle$, at least in the range $\ell\leq O(\eta_B)$. This then implies that the inverse buoyancy flux at the small scales should become stronger as $Pr$ increases. This is consistent with what has been previously observed in DNS of stratified turbulence with $Pr\geq 1$ \citep{Okino2019,Legaspi2020}.

\section{Direct Numerical Simulations}\label{sec:Numerical}

The data sets to be used in the next section are from the simulation campaign of statistically stationary, stably-stratified DNSs first described in \cite{AlmalkiedBK2012}, which have later been examined by \citet{debk15}, \cite{Portwood2016}, \cite{Taylor2019}, \cite{CouchmandBKCC2023} and \cite{Petropoulos2024}. The DNS are for $Pr=1,7$ and $Fr\approx 0.08, 0.16$ (note that the definition of $Fr$ differs by a factor of $2\pi$ relative to the references cited at the start of this section).  In each case, a constant activity parameter \textcolor{black}{(which is closely related to the buoyancy Reynolds number defined in terms of the integral scales of the flow; see \citet{deBKRiley_2019} for a discussion)} $Gn \equiv \langle\epsilon \rangle/(\nu N^2) \approx 50$ is maintained. Table \ref{tab:Parameters} provides a summary of the parameters from the current DNS; additional details of the simulation algorithm can be found in \cite{AlmalkiedBK2012}.

For the post-processing analysis, an isotropic Gaussian filter was used, consistent with the theory in \S \ref{sec:Theory}. Logarithmically spaced filter-scales were used and the set of $\ell/\eta$ values used in the analysis is the same for each case.

\begin{table}
\centering
\captionsetup{width=\linewidth}
\begin{tabular}{c c c c c c c c c} 
$Gn$    &   $Pr$    &   $Fr$     &  AR   &   $n_h$   &   $L/\eta$  & $L/l_{Oz}$&   $l_{Oz}/\eta$ &   $\eta/\eta_B$  \\ \\ 
$50$    &   $1$     &   $0.08$   &  $8$  &   $8192$  &   $1686.8$  & $86.1$    &   $19.6$       &   $1$             \\ 
$50$    &   $1$     &   $0.16$   &  $4$  &   $4096$  &   $649.8$   & $27.6$    &   $23.6$       &   $1$             \\ 
$50$    &   $7$     &   $0.08$   &  $8$  &   $30240$ &   $1362.0$  & $66.0$    &   $20.6$       &   $\sqrt{7}$      \\ 
$50$    &   $7$     &   $0.16$   &  $4$  &   $8192$  &   $511.4$   & $27.2$    &   $18.8$       &   $\sqrt{7}$      \\ 
\end{tabular}
\caption{Flow parameters in DNS. $Gn\equiv \langle \epsilon \rangle/\nu N^2$ is the activity parameter, $Fr\equiv U/(LN)$ is the Froude number. AR$=\mathcal{L}_h/\mathcal{L}_z$ is the aspect ratio of the simulation domain, where $\mathcal{L}_h=2 \pi$, $\mathcal{L}_z$ are the lengths of the domain in the horizontal ($h$) and the vertical ($z$) directions, respectively. 
\textcolor{black}{The 3D simulation domain is discretized with ($n_h$,$n_h$,$n_z$) grids points, such that $n_z=n_h/\text{AR}$.} 
$L$ is the integral length scale computed as the horizontal autocorrelation length of the horizontal field. $l_{Oz}$, $\eta$ and $\eta_B$ are the Ozmidov, Kolmogorov and Bachelor length scales, respectively. 
\label{tab:Parameters}}
\end{table}


\section{Results \& Discussion}\label{sec:Results}

\subsection{Sign of $\langle\mathcal{P}^\ell_{B2}\rangle$ across scales}

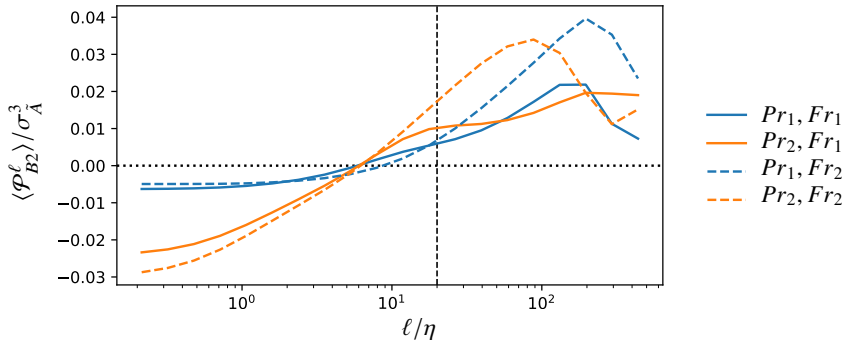
\begin{figure}
\centering
\includegraphics[trim = {0mm 6mm 0mm 0mm}, scale=0.6,clip]{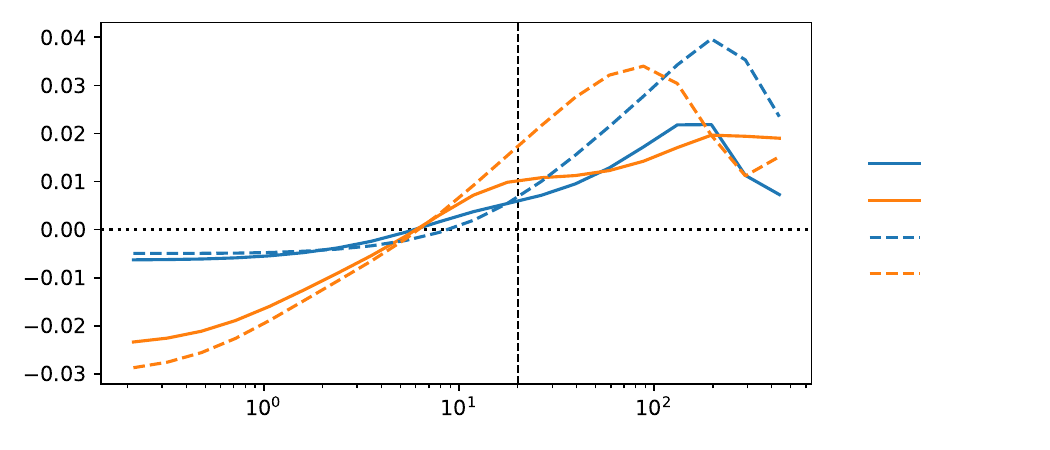}
\put(-315,40){\rotatebox{90}{{\footnotesize $\langle\mathcal{P}^\ell_{B2}\rangle/\sigma^3_{\tilde{A}}$}}}
\put(-32,70.0){{\footnotesize $Pr_1, Fr_1$}}
\put(-32,60.0){{\footnotesize $Pr_2, Fr_1$}}
\put(-32,50.0){{\footnotesize $Pr_1, Fr_2$}}
\put(-32,40.0){{\footnotesize $Pr_2, Fr_2$}}
\put(-167,-10){{\footnotesize $\ell/\eta$}}
\caption{Lin-log plots of $\langle\mathcal{P}^\ell_{B2}\rangle/\sigma^3_{\tilde{A}}$ as a function of scale for $Pr_1=1$ (blue) and $Pr_2=7$ (orange) with $Fr_1 \approx 0.08$ (solid) and $Fr_2 \approx 0.16$ (dashed). }
\label{fig:PB2l_Preffect}
\end{figure}

We first examine the behaviour of the filtered production mechanism $\langle \mathcal{P}^\ell_{B2}\rangle$, which is associated with the local contribution to the scale-dependent buoyancy flux $\langle\mathcal{B}\rangle$. 
Figure \ref{fig:PB2l_Preffect} shows $\langle\mathcal{P}^\ell_{B2}\rangle/\sigma^3_{\tilde{A}}$ (where $\sigma_{\tilde{A}}\equiv \langle\| \tilde{\bA}\|^2\rangle^{1/2}$) as a function of $\ell/\eta$ for $Fr_1 \approx 0.08$ {(solid lines)} and $Fr_2 \approx 0.16$ {(dashed lines)} and for $Pr=1$ (blue), and $Pr=7$ (orange). The results clearly show that $\langle\mathcal{P}^\ell_{B2}\rangle$ changes sign as $\ell$ is decreased, in agreement with the arguments in \cite{BraggdBKops_2024}. 
The DNS data in \cite{BraggdBKops_2024} had also shown this sign change, but their data showed that $\langle\mathcal{P}^\ell_{B2}\rangle$ switched sign again and became negative at the largest scales of the flow. They suggested that this was due to the non-stationarity of the decaying stratified turbulent flow they considered, since for a stationary flow their analysis suggested that $\langle\mathcal{P}^\ell_{B2}\rangle$ must be positive in order to satisfy the balance equation for $\langle\|\tilde{\bm B}\|^2\rangle$ at the large scales \textcolor{black}{\eqref{eq:filteredchi_budget}}. The present results which are for a stationary stratified flow seem to confirm this, showing that $\langle\mathcal{P}^\ell_{B2}\rangle$ remains positive at the largest flow scales.

\subsection{\textcolor{black}{Reversal of average b}uoyancy flux}

\begin{figure}
\centering
\includegraphics[trim = {0mm 2mm 0mm 0mm}, scale=0.55,clip]{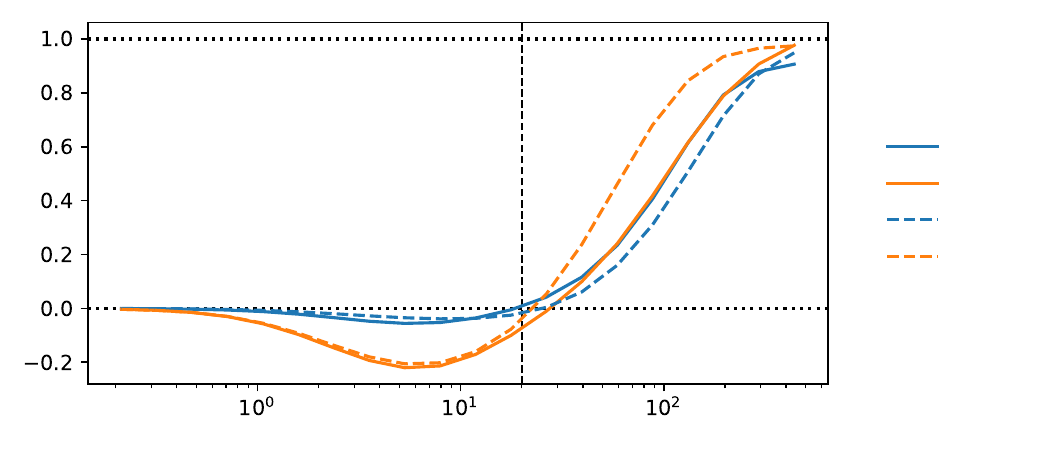}
\put(-32,80.0){{\footnotesize $Pr_1, Fr_1$}}
\put(-32,70.0){{\footnotesize $Pr_2, Fr_1$}}
\put(-32,60.0){{\footnotesize $Pr_1, Fr_2$}}
\put(-32,50.0){{\footnotesize $Pr_2, Fr_2$}}
\put(-167,-10){{\footnotesize $\ell/\eta$}}
\put(-285,50){\rotatebox{90}{{\footnotesize $-\langle\mathcal{B}\rangle/\langle \chi \rangle$}}}
\caption{Buoyancy flux $\langle \mathcal{B} \rangle$ as a function of scale $\ell$ for $Pr=1$ (blue) and $Pr=7$ (orange) with $Fr_1 \approx 0.08$ (solid) and $Fr_2 \approx 0.16$ (dashed).
Black vertical dashed line marks the approximate Ozmidov scale $l_{O}/\eta \sim 20$, exact values in table \ref{tab:Parameters}.}
\label{fig:BuoyancyFluxPr}
\end{figure}
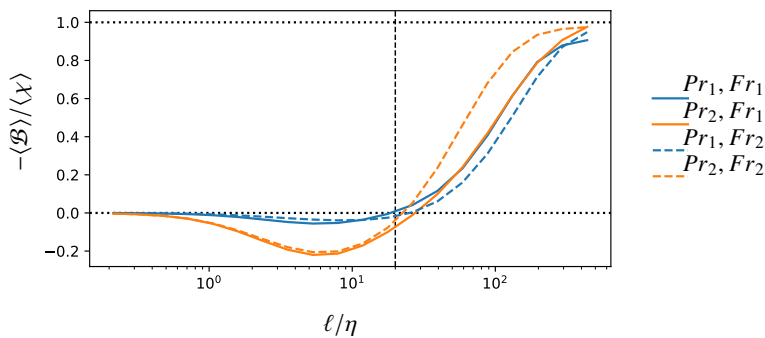

Figure \ref{fig:BuoyancyFluxPr} shows plots of $-\langle\mathcal{B}\rangle/\langle\chi\rangle$ versus $\ell/\eta$ for $Pr_1=1$ (blue) and $Pr_2=7$ (orange) for $Fr_1 \approx 0.08$ (solid) and $Fr_2 \approx 0.16$ (dashed). At the large scales $-\langle \mathcal{B} \rangle>0$ and $-\langle \mathcal{B} \rangle\sim \langle \chi \rangle$, reflecting a balance between the rate of production of TPE due to buoyancy and the dissipation rate of TPE. However, for $\ell \lesssim l_{O}$ the results show that $-\langle \mathcal{B} \rangle$ switches sign and becomes negative, indicating a reverse buoyancy flux with TPE being converted back into TKE at the small scales. This flux reversal occurs for both $Pr=1$ and $Pr=7$, but the magnitude of the reverse flux is much more significant for $Pr=7$. \cite{Legaspi2020} previously observed this flux-reversal in their DNS results when analyzed in Fourier space, and they argued that the inverse buoyancy flux is the reason why simulations with $Pr>1$ show shallowing of the TKE spectra at high wave-numbers (see Fig. 10 of \cite{Legaspi2020}, and Fig. 8 of \cite{RileyCdBKops_2023}. \cite{Legaspi2020} observed a stronger buoyancy flux reversal with decreasing $Fr$, which we also see here (although the effect of $Fr$ is very weak for the value of $Gn$ in our DNS), and also observed that the magnitude of the reverse buoyancy flux increased with increasing $Pr$, which we again see here. They observed a saturation in the $Pr$ dependence of the buoyancy flux spectra as $Pr$ was increased, while \citet{Okino2019} found that the buoyancy flux reversal continues to grow in magnitude when going from $Pr=7$ to $Pr=70$. It is likely that any saturation of the effect of $Pr$ occurs at a value of $Pr$ that depends on both the buoyancy Reynolds number $Re_b$ and $Fr$, and these values were different in \cite{Legaspi2020} and \citet{Okino2019}.

\subsection{Contribution to average buoyancy flux from local and non-local mechanisms} 

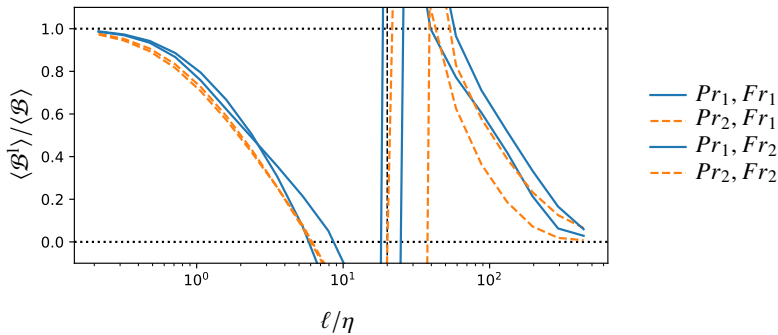
\begin{figure}
\centering
\includegraphics[trim = {0mm 2mm 0mm 0mm}, scale=0.55,clip]{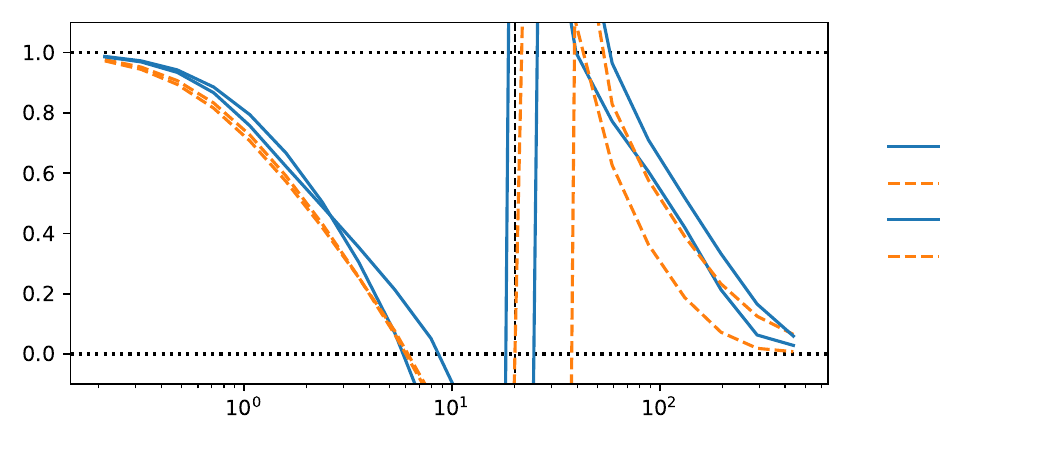}
\put(-285,45){\rotatebox{90}{{\footnotesize $\langle \mathcal{B}^{{\rm l}}\rangle/ \langle\mathcal{B}\rangle$}}} %
\put(-26,75.0){{\footnotesize $Pr_1, Fr_1$}}
\put(-26,66.0){{\footnotesize $Pr_2, Fr_1$}}
\put(-26,56.0){{\footnotesize $Pr_1, Fr_2$}}
\put(-26,46.0){{\footnotesize $Pr_2, Fr_2$}}
\put(-167,-10){{\footnotesize $\ell/\eta$}}
\caption{Ratio of the scale-local component to the total buoyancy flux $\langle \mathcal{B}^{{\rm l}}\rangle/\langle\mathcal{B}\rangle$ as a function of filter-scale $\ell$ for $Pr_1=1$ (blue), $Pr_2=7$ (orange) with $Fr_1 \approx 0.08$ (solid) and $Fr_2 \approx 0.16$ (dashed).
Black vertical dashed line marks the approximate Ozmidov scale $l_{O}/\eta \sim 20$, exact values in table \ref{tab:Parameters}.
}
\label{fig:LocaltotalB}
\end{figure}

As discussed in \S\ref{sec:Theory}, the filtered production mechanism $ \mathcal{P}^{\ell}_{B2}$ is related to the scale-local part of the buoyancy flux $\textcolor{black}{\mathcal{B}^{{\rm l}}}=-\ell^2  \mathcal{P}^\ell_{B2}$, i.e. the part of the buoyancy flux involving only contributions from filtered scales of motion. 
Figure \ref{fig:LocaltotalB} shows the ratio of $\textcolor{black}{\langle\mathcal{B}^{{\rm l}}\rangle}$ to $\langle\mathcal{B}\rangle$ for $Pr_1=1$ (blue), $Pr_2=7$ (orange) with $Fr_1 \approx 0.08$ (solid) and $Fr_2 \approx 0.16$ (dashed). 
In the limit $\ell/\eta\to 0$ the ratio approaches one, consistent with the analysis in \S\ref{sec:Theory} that predicts $\lim_{\ell/\eta\to 0}\mathcal{B}\to -\ell^2\mathcal{P}_{B2}$. 
While there are extended portions of the scale space over which $\langle\mathcal{B}^{{\rm l}}\rangle$ makes a substantial contribution to $\langle\mathcal{B}\rangle$, it is clear that in general the sub-\textcolor{black}{filter} term $\langle\mathcal{B}^{{\rm nl}}\rangle$ makes a significant contribution. 
At intermediate $\ell/\eta$, $\langle\mathcal{B}^{{\rm l}}\rangle/\langle\mathcal{B}\rangle$ diverges to $+\infty$, and then to $-\infty$, and this occurs because $\langle \mathcal{P}^\ell_{B2}\rangle$ and $\langle\mathcal{B}\rangle$ do not change sign at the same scale. 
Outside of the range of $\ell/\eta$ in which these divergences occur, the signs of $\langle\mathcal{B}^{{\rm l}}\rangle$ and $\langle\mathcal{B}\rangle$ correspond to each other. 
The results also show that in the limit $\ell/\eta\to \infty$, $\langle\mathcal{B}^{{\rm l}}\rangle/\langle\mathcal{B}\rangle\to 0$, consistent with the argument made earlier that in this limit $\langle\mathcal{B}\rangle\to\langle\mathcal{B}^{{\rm nl}}\rangle$ due to homogeneity of the flow.  

Traditionally, the small-scale reverse buoyancy flux has been related to internal wave breaking and convective instabilities \citep{Holloway1988,BA_1996}. \cite{Holloway1988} proposed that the TPE is transferred more efficiently to smaller scales than TKE through the cascade mechanism, and therefore the reverse buoyancy flux restores equipartition. \cite{BA_1996} argued that it is due to shear-instabilities in vorticity layers that form after wave-breaking. While these explanations are possible, the scale at which the reversal occurs seems too small for the underlying mechanism to be associated with wave motion which is more likely to play a role at scales $\ell>l_{O}$. There is also no reason why equipartition should be expected to hold for these non-equilibrium systems. More importantly, however, is that these mechanisms  cannot explain why $\langle\mathcal{B}\rangle$ also changes sign at the small-scales even for a passive scalar (neutrally buoyant) flow (for such a flow, although $\langle\mathcal{B}\rangle$ is absent from the TKE equation, it is still present in the TPE equation and describes the production of $\phi$ due to the mean scalar gradient). As discussed earlier, at the small scales $\langle\mathcal{B}\rangle\sim -\ell^2\langle \mathcal{P}_{B2}\rangle$, and \cite{BraggdBKops_2024} showed that $\langle \mathcal{P}_{B2}\rangle$ is negative for a passive scalar due to the formation of ramp-cliffs, implying $\langle\mathcal{B}\rangle>0$ at those scales. Since the mechanism we have proposed, which is grounded in that described by \cite{BraggdBKops_2024}, can explain why $\langle\mathcal{B}\rangle>0$ at the small-scales in both stratified and neutral flows, it seems more plausible than the alternative explanations which could only apply to stratified flows. However, while the mechanism we have proposed explains why $\langle\mathcal{B}\rangle>0$ at the smallest scales, it only partially explains it outside of this range. A complete explanation would require a mechanism that also explains why $\langle\mathcal{B}^{{\rm nl}}\rangle>0$ at smaller scales. 

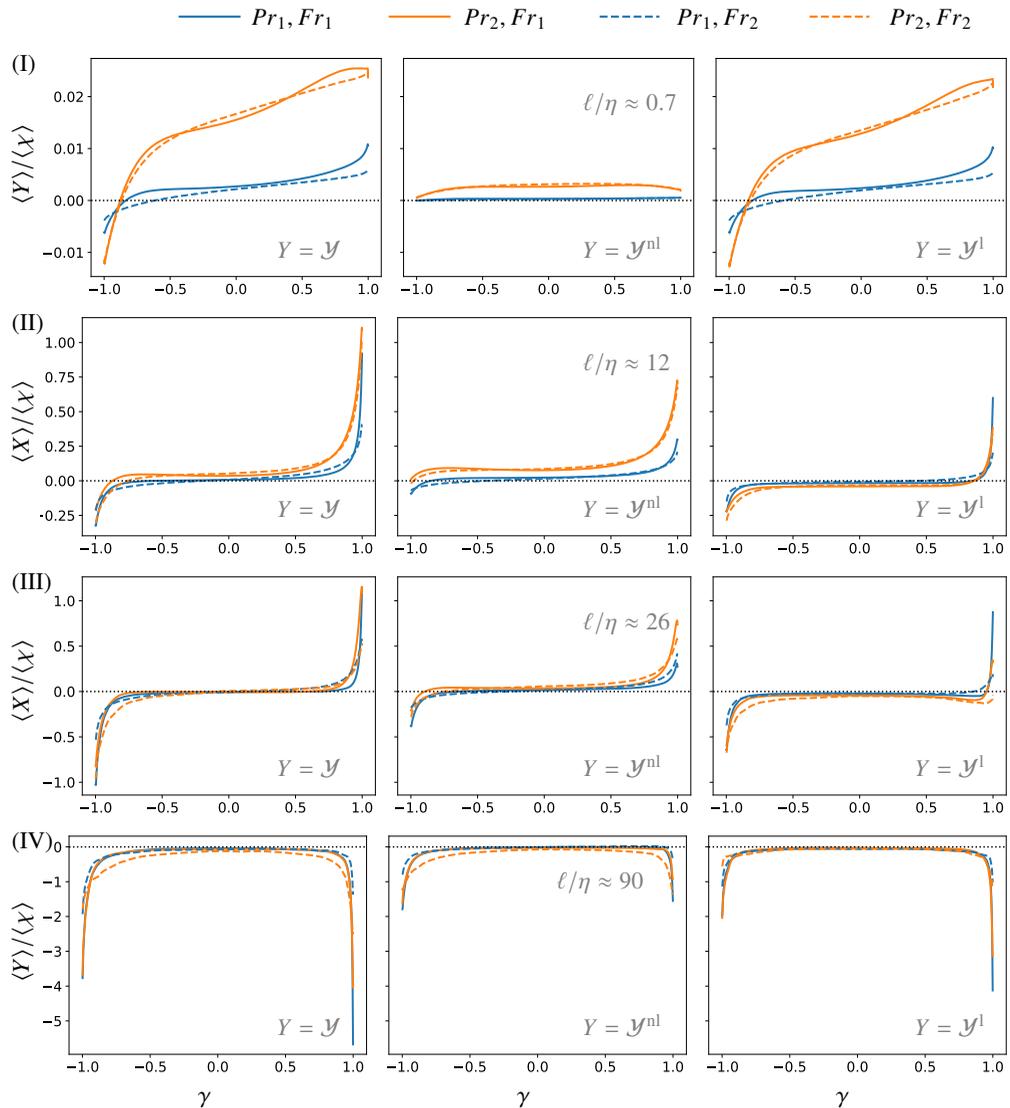
\begin{figure}
\centering
\includegraphics[trim = {0mm 0mm 20mm 10mm}, scale=0.4,clip]{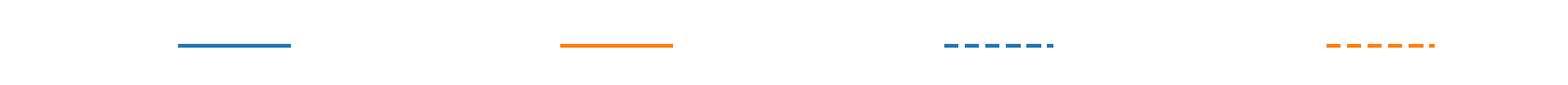}
\put(-260,5.0){{\footnotesize $Pr_1, Fr_1$}}
\put(-180,5.0){{\footnotesize $Pr_2, Fr_1$}}
\put(-100,5.0){{\footnotesize $Pr_1, Fr_2$}}
\put(-20, 5.0){{\footnotesize $Pr_2, Fr_2$}}

\flushright
\includegraphics[trim = {5mm 13mm 0mm 10mm}, scale=0.37,clip]{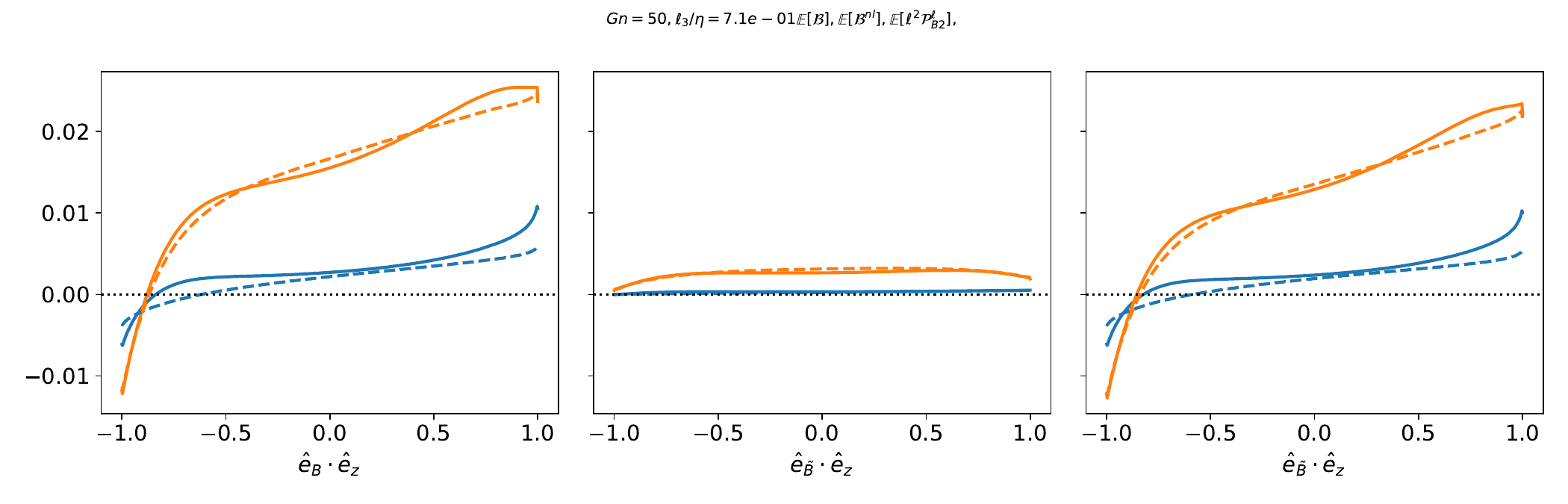}
\put(-380,35){\rotatebox{90}{{\footnotesize $\langle Y \rangle/\langle \chi \rangle$}}}
\put(-380,85){{\footnotesize (I)}}
\put(-165,70){\textcolor{gray}{\footnotesize $\ell/\eta \approx 0.7$}}
\put(-280,15){\textcolor{gray}{\footnotesize $Y=\mathcal{Y}$}}
\put(-165,15){\textcolor{gray}{\footnotesize $Y=\mathcal{Y}^{{\rm nl}}$}}
\put( -40,15){\textcolor{gray}{\footnotesize $Y=\mathcal{Y}^{{\rm l}}$}}


\includegraphics[trim = {5mm 13mm 0mm 10mm}, scale=0.37,clip]{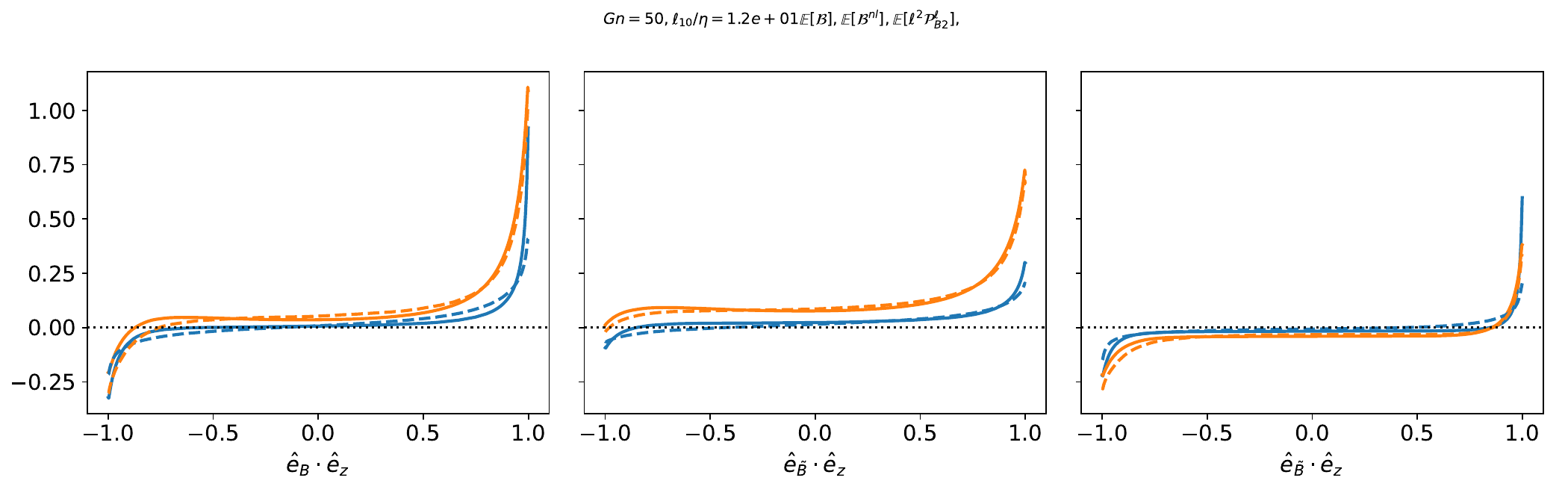}
\put(-380,35){\rotatebox{90}{{\footnotesize $\langle X \rangle/\langle \chi \rangle$}}}
\put(-380,85){{\footnotesize (II)}}
\put(-165,70){\textcolor{gray}{\footnotesize $\ell/\eta \approx 12$}}
\put(-280,15){\textcolor{gray}{\footnotesize $Y=\mathcal{Y}$}}
\put(-165,15){\textcolor{gray}{\footnotesize $Y=\mathcal{Y}^{{\rm nl}}$}}
\put( -40,15){\textcolor{gray}{\footnotesize $Y=\mathcal{Y}^{{\rm l}}$}}

\includegraphics[trim = {5mm 13mm 0mm 10mm}, scale=0.37,clip]{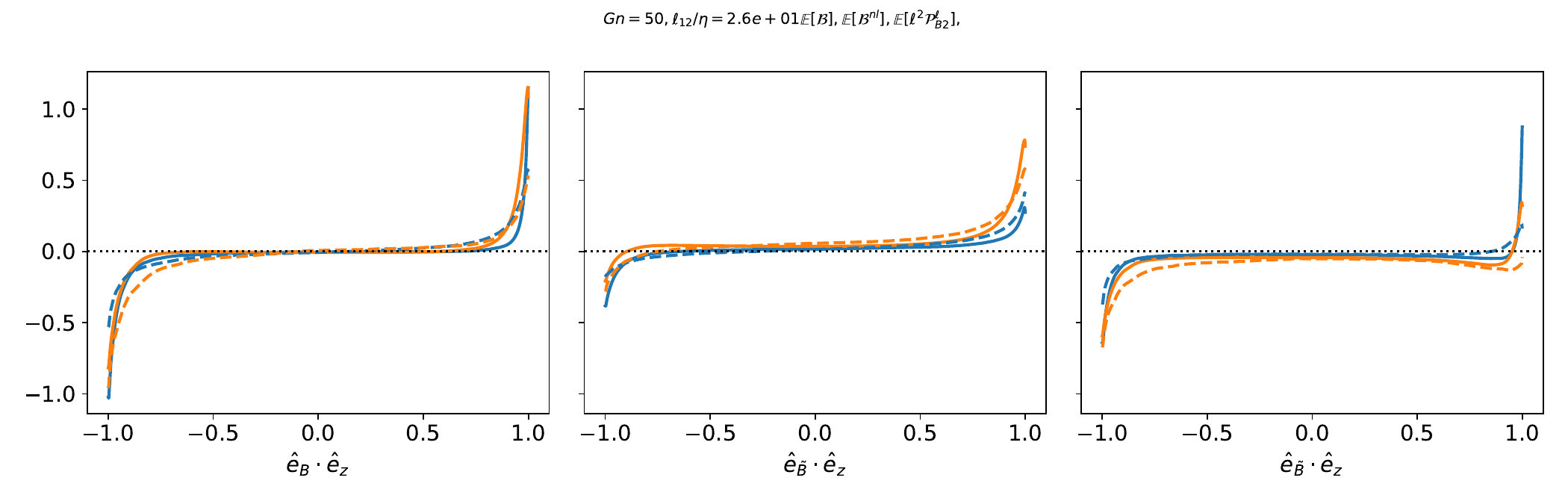}
\put(-380,35){\rotatebox{90}{{\footnotesize $\langle X \rangle/\langle \chi \rangle$}}}
\put(-380,85){{\footnotesize (III)}}
\put(-165,70){\textcolor{gray}{\footnotesize $\ell/\eta \approx 26$}}
\put(-280,15){\textcolor{gray}{\footnotesize $Y=\mathcal{Y}$}}
\put(-165,15){\textcolor{gray}{\footnotesize $Y=\mathcal{Y}^{{\rm nl}}$}}
\put( -40,15){\textcolor{gray}{\footnotesize $Y=\mathcal{Y}^{{\rm l}}$}}

\includegraphics[trim = {5mm 13mm 0mm 10mm}, scale=0.37,clip]{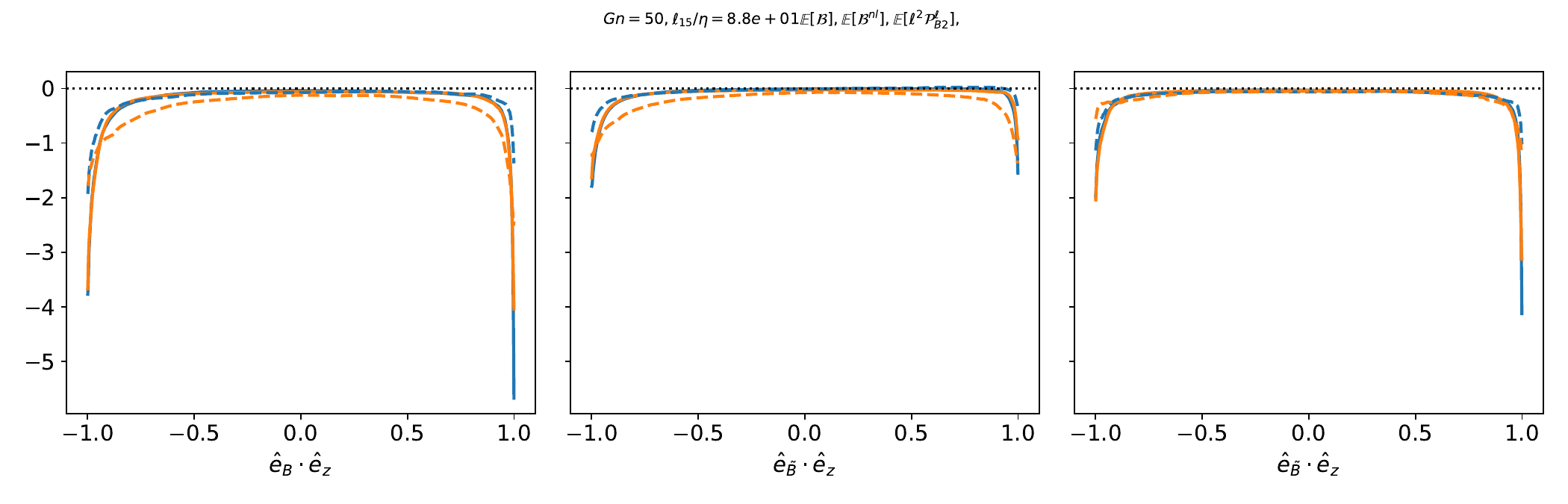}
\put(-380,35){\rotatebox{90}{{\footnotesize $\langle Y \rangle/\langle \chi \rangle$}}}
\put(-380,85){{\footnotesize (IV)}}
\put(-175,70){\textcolor{gray}{\footnotesize $\ell/\eta \approx 90$}}
\put(-280,15){\textcolor{gray}{\footnotesize $Y=\mathcal{Y}$}}
\put(-165,15){\textcolor{gray}{\footnotesize $Y=\mathcal{Y}^{{\rm nl}}$}}
\put( -40,15){\textcolor{gray}{\footnotesize $Y=\mathcal{Y}^{{\rm l}}$}}
\put(-310,-10){{\footnotesize $\gamma $}}
\put(-190,-10){{\footnotesize $\gamma $}}
\put( -70,-10) {{\footnotesize $\gamma $}}

\caption{\textcolor{black}{Averages of the total buoyancy flux $\mathcal{B}$, and its scale-local and non-local  decompositions $\mathcal{B}^{{\rm l}}$ and $\mathcal{B}^{{\rm nl}}$, respectively, conditioned on $\gamma=\hat{\bm{e}}_{\tilde{B}} \bm{\cdot} \ez$ and normalised by the average TPE dissipation rate $\langle\chi\rangle$. Blue (orange) lines show the values for $Pr_1=1$ ($Pr_2=7$), whereas the solid (dashed) lines correspond to $Fr_1\approx0.08$ and $Fr_2\approx 0.16$, respectively.
Panels I, II, III, and IV correspond to fields filtered at scales $\ell/\eta \approx 0.7$, $\ell/\eta \approx 12$, $\ell/\eta \approx 26$, and $\ell/\eta \approx 90$, respectively.
\label{fig:BBnlBl_tBz}}}  %
\end{figure}

\textcolor{black}{An important observation is that the integrand defining the non-local buoyancy flux $\mathcal{B}^{{\rm nl}}$ (see equation \eqref{eq:partition_PB}) is mathematically similar in form to the quantity $\tilde{\bB} \bm{\cdot} \tilde{\bA}^\top \bm{\cdot} \ez$ appearing in the definition of $\mathcal{P}^\ell_{B2}$, with both involving invariants formed from filtered velocity and density gradients, along with $\ez $. 
This suggests that just as the sign of $\mathcal{P}^\ell_{B2}$ depends crucially on the sign of $\hat{\bm{e}}_{\tilde{B}} \bm{\cdot} \ez $ (where $\hat{\bm{e}}_{\tilde{B}} \equiv \widetilde{\bm B}/\|\widetilde{\bm B}\|$) and the associated ramp-cliff structures, so also might $\mathcal{B}^{{\rm nl}}$. 
If so, then the formation of ramp-cliff structures might not only be responsible for $\langle\mathcal{P}^\ell_{B2}\rangle$ becoming negative at small scales, but also for $\langle\mathcal{B}^{{\rm nl}}\rangle$ and hence the entire flux $\langle\mathcal{B}\rangle$ becoming positive at small scales. 
To explore this further, we conditionally average the buoyancy flux and its local and non-local components based on the alignment of the density gradient $\hat{\bm{e}}_{\tilde{B}} \bm{\cdot} \ez \in[-1,+1]$ at different filter scales, defined as
\begin{align}
\mathcal{Y}(\gamma)&\equiv \langle \mathcal{B}\rangle_\gamma\mathcal{P}(\gamma),\\    
\mathcal{Y}^{{\rm nl}}(\gamma)&\equiv \langle \mathcal{B}^{{\rm nl}}\rangle_\gamma\mathcal{P}(\gamma),\\    
\mathcal{Y}^{{\rm l}}(\gamma)&\equiv \langle \mathcal{B}^{{\rm l}}\rangle_\gamma\mathcal{P}(\gamma),
\end{align}
where $\langle\cdot\rangle_\gamma$ denotes an average conditioned on $\gamma=\hat{\bm{e}}_{\tilde{B}} \bm{\cdot} \ez$, $\mathcal{P}(\gamma)\equiv\langle\delta( \gamma- \hat{\bm{e}}_{\tilde{B}} \bm{\cdot} \ez )\rangle$ is the probability density function of $\gamma$, and $\langle \mathcal{B}\rangle=\int_{-1}^{+1}\mathcal{Y}(\gamma)\,d\gamma$, $\langle \mathcal{B}^{{\rm nl}}\rangle=\int_{-1}^{+1}\mathcal{Y}^{{\rm nl}}(\gamma)\,d\gamma$, $\langle \mathcal{B}^{{\rm l}}\rangle=\int_{-1}^{+1}\mathcal{Y}^{{\rm l}}(\gamma)\,d\gamma$. 
Ramps (which are the most frequently occurring events) correspond to $\gamma > 0$ (i.e. misalignment of $\tilde{\bB}$ with the mean density direction $\hat{\bg}=-\ez$), whereas cliffs correspond to  $\gamma < 0$.
Figure \ref{fig:BBnlBl_tBz} shows the results for $\mathcal{Y},\mathcal{Y}^{{\rm nl}}, \mathcal{Y}^{{\rm l}}$ at different scales, where blue (orange) lines show the values for $Pr_1=1$ ($Pr_2=7$), whereas the solid (dashed) lines correspond to $Fr_1\approx0.08$ ($Fr_2\approx 0.16$).
Panels I, II, III, and IV correspond to filtering at scales $\ell/\eta \approx 0.7$, $\ell/\eta \approx 12$, $\ell/\eta \approx 26$, and $\ell/\eta \approx 90$, respectively.}

\textcolor{black}{The results for $\mathcal{Y}$ show that at the small scales where there is a reverse buoyancy flux (i.e. $\ell/\eta\lesssim 26$), ramp structures contribute exclusively to reverse buoyancy flux events, and at intermediate scales (e.g. $\ell/\eta \approx 12, 26$) the main contribution is from ramp regions with strong alignment $\gamma \approx +1$. 
By contrast, cliff regions contribute to both forward and reverse buoyancy flux events, with regions of strong misalignment $\gamma\approx -1$ strikingly associated with regions of forward buoyancy flux at all scales in the flow, where TKE is converted to TPE. 
These results are consistent with the arguments of \cite{BraggdBKops_2024} and those presented in \S\ref{sec:Theory}, lending support to the idea that the sign of the buoyancy flux is intimately connected to ramp-cliff structures in the flow. 
It is also quite remarkable that except for scales $\ell\leq O(\eta)$, the largest contributions to $\langle\mathcal{B}\rangle$ come from the extreme events where $\gamma\approx -1$ and $\gamma\approx +1$, further emphasizing the crucial role that ramp-cliffs play in the buoyancy flux mechanism.}

\textcolor{black}{The results for $\mathcal{Y}^{{\rm nl}}$ are qualitatively very similar to those of both $\mathcal{Y}$ and $\mathcal{Y}^{{\rm l}}$ (except for $\ell/\eta\approx 0.7$, which is simply because $\mathcal{Y}^{{\rm nl}}\to 0$ for $\ell/\eta\to 0$). 
In particular, $\mathcal{Y}^{{\rm nl}}$ exhibits a reverse flux in ramp regions where $\gamma \approx +1$, and a forward flux in cliff regions where $\gamma \approx -1$. 
This strongly suggests that the formation of ramp-cliff structures is not only responsible for the behaviour of the local part of the buoyancy flux and its sign, but also that of the non-local part, as speculated earlier. 
At larger scales,  $\ell/\eta \approx 90$, $\mathcal{Y}, \mathcal{Y}^{{\rm nl}}, \mathcal{Y}^{{\rm l}}$ are negative at all $\gamma$, corresponding to a total buoyancy flux that is forward, converting TKE to TPE. 
For a statistically stationary, stably-stratified turbulent flow where only the TKE field has a source term (as in our DNS), the TPE field cannot be sustained unless $\langle\mathcal{B}\rangle<0$ at the large scales. 
There are therefore two effects controlling the sign of $\langle\mathcal{B}\rangle$, one coming from the symmetry-breaking effect of the ramp-cliff structures, and the other from the need to sustain the TPE field. 
In a sense the latter is more fundamental because unless the TPE field is sustained, there would not even be ramp-cliff structures in the first place. 
We therefore suggest the following picture for understanding the behaviour of the sign of $\langle\mathcal{B}\rangle$: At large scales, $\langle\mathcal{B}\rangle<0$ which must occur in order for the TPE field to be sustained against the dissipative effect of molecular mixing of the density field. 
However, once sufficiently small scales are attained for $\langle\mathcal{B}\rangle$ to play a sub-leading role in the TPE equation (e.g. because the TPE interscale flux term becomes important) then the symmetry-breaking effect of the ramp-cliffs is able to exert its influence and enforce $\langle\mathcal{B}\rangle>0$.}


\begin{figure}
\centering
\includegraphics[trim = {0mm 2mm 0mm 0mm}, scale=0.57,clip]{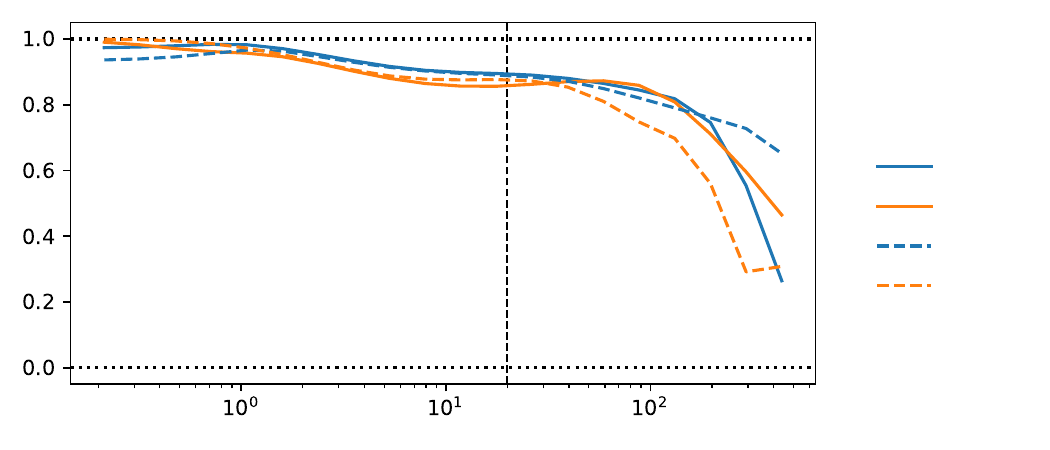}
\put(-295,35){\rotatebox{90}{{\footnotesize Corr$(\mathcal{B},\mathcal{B}^{{\rm l}})$}}}
\put(-30,72.0){{\footnotesize $Pr_1, Fr_1$}}
\put(-30,61.0){{\footnotesize $Pr_2, Fr_1$}}
\put(-30,50.0){{\footnotesize $Pr_1, Fr_2$}}
\put(-30,40.0){{\footnotesize $Pr_2, Fr_2$}}
\put(-167,-10){{\footnotesize $\ell/\eta$}}
\caption{Correlation coefficient of $\mathcal{B}$
with the scale local term $\mathcal{B}^{{\rm l}}=-\ell^2 \mathcal{P}^\ell_{B2}$ 
(Eq.\ref{eq:partition_PB}) for $Pr_1=1$ (blue), $Pr_2=7$ (orange) with $Fr_1 \approx 0.08$ (solid) and $Fr_2 \approx 0.16$ (dashed). 
Black vertical dashed line marks the approximate Ozmidov scale $l_{O}/\eta \sim 20$, exact values in table \ref{tab:Parameters}.
\label{fig:Corr_BPB2}}  %
\end{figure}
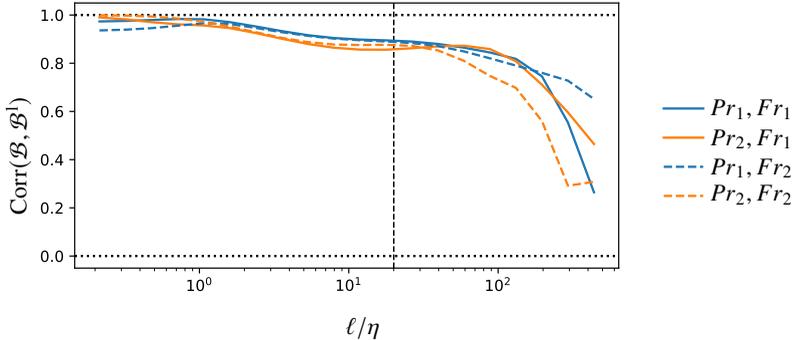

\subsection{Contribution from local mechanism to fluctuations of buoyancy flux} 

Finally, we turn to consider how fluctuations of the local contribution $\mathcal{B}^{{\rm l}}$ are correlated with those of the instantaneous buoyancy flux $\mathcal{B}$. 
Figure \ref{fig:Corr_BPB2} shows the correlation coefficient of $\mathcal{B}^{{\rm l}}$ and $\mathcal{B}$ as a function of $\ell/\eta$ for $Pr_1=1$ (blue), $Pr_2=7$ (orange) with $Fr_1 \approx 0.08$ (solid) and $Fr_2 \approx 0.16$ (dashed). 
We observe that the total buoyancy flux $\mathcal{B}$ is strongly correlated to the local contribution $\mathcal{B}^{{\rm l}}$, and this holds true even at scales where the ratios of their mean values is significantly different. 
\textcolor{black}{The strong correlation in the fluctuations of $\mathcal{B}$ and $\mathcal{B}^{{\rm l}}$, especially at scales when $\mathcal{B}^{{\rm nl}}$ is significant, is further evidence of the shared underlying mechanism, related to ramp-cliff structures, shown in Fig. \ref{fig:BBnlBl_tBz}.}
This is reminiscent of the correlation observed in isotropic turbulence between the TKE inter-scale energy flux $\Pi_K$ and the scale-local contribution to this flux involving the combined effects of strain self-amplification $\Pi_K^{SSA}$ and vortex stretching $\Pi_K^{VS}$. In particular, \citet{Johnson_2021} showed that while $\langle \Pi_K^{SSA}+\Pi_K^{VS}\rangle/\langle\Pi_K\rangle\approx 0.48$ in the inertial range, the correlation coefficient between $\Pi_K$ and $\Pi_K^{SSA}+\Pi_K^{VS}$ has a value close to $0.9$ in the same range, which is also what we observe here. The results in Fig. \ref{fig:Corr_BPB2} also indicate that $Fr,Pr$ do not have a strong or systematic effect on the correlations, except at scales $\ell/\eta\geq O(100)$.

\section{Conclusions}\label{sec:Conclusions}
Recent investigations of stratified turbulence show that the TPE (TKE) dissipation rate  $\langle\chi\rangle$ ($\langle\epsilon\rangle$) is smaller (larger) for flows with larger $Pr$ \citep{Okino2019,RileyCdBKops_2023}.
\cite{BraggdBKops_2024} argued that this can be understood due to the role of the gradient production mechanism associated with the mean density gradient, $\mathcal{P}_{B2}$, which appears with an opposite sign in the velocity and density gradient equations. In particular they argued that the formation of ramp-cliff structures in the fluctuating density field ultimately leads to $\langle\mathcal{P}_{B2}\rangle$ being negative, and the fact that the fluctuating density gradients increase in magnitude as $Pr$ is increased also causes the magnitude of  $\langle\mathcal{P}_{B2}\rangle$ to increase. This then causes $\langle\chi\rangle$ to decrease and $\langle\epsilon\rangle$ to increase as $Pr$ increases. In this study we sought to explain the connection between this mechanism and the small-scale reverse buoyancy flux that has previously been observed in stratified turbulence.

An exact decomposition of the scale-dependent buoyancy flux $\mathcal{B}$ was derived using a filtering analysis and analytical results from \cite{Johnson_2020,Johnson_2021}. The result is $\mathcal{B}=-\ell^2\mathcal{P}_{B2}^\ell+\mathcal{B}^{SG}$, where $\mathcal{P}_{B2}^\ell$ is a generalization of the gradient production term analysed in \cite{BraggdBKops_2024} that describes the production of gradients filtered at scale $\ell$ and satisfies $\mathcal{P}_{B2}^\ell=\mathcal{P}_{B2}$ for $\ell/\eta_B\to 0$ (where $\eta_B$ is the Batchelor length). The contribution $-\ell^2\mathcal{P}_{B2}^\ell$ is the scale-local contribution representing the contribution from scales $\geq\ell$, while $\mathcal{B}^{SG}$ is the sub-grid or non-local contribution. In the limit $\ell/\eta_B\to 0$ the result yields $\langle\mathcal{B}\rangle=-\ell^2\langle\mathcal{P}_{B2}\rangle$, and therefore the mechanism presented in \cite{BraggdBKops_2024} for why $\langle\mathcal{P}_{B2}\rangle<0$ and why its magnitude increases with increasing $Pr$ also explains why $\langle\mathcal{B}\rangle>0$ at the smallest scales of stratified turbulence and why the reverse buoyancy flux at the small scales becomes stronger with increasing $Pr$. 

State-of-the-art DNS \textcolor{black}{of three-dimensional, }strongly stratified turbulence with $Pr=1,7$ was used to test the predictions and explore further the behaviour of $\mathcal{B}$. 
The results confirm the argument from \cite{BraggdBKops_2024} that $\langle\mathcal{P}_{B2}^\ell\rangle$ should switch from being positive at the large scales, to negative at the small scales. 
The results for $\langle\mathcal{B}\rangle$ also agree with previous studies showing that $\langle\mathcal{B}\rangle>0$ at the small-scales, with a magnitude that increases with increasing $Pr$. 
The results confirm that for $\ell\leq O(\eta_B)$, $\langle\mathcal{B}\rangle\sim-\ell^2\langle\mathcal{P}_{B2}\rangle$, and therefore support the proposed mechanism for why $\langle\mathcal{B}\rangle>0$ at the smallest scales. 
Although $\langle\mathcal{B}^{{\rm nl}}\rangle$ makes an important contribution to $\langle\mathcal{B}\rangle$ \textcolor{black}{beyond the smallest scales,}
we show that the correlation between the scale-local part $-\ell^2\langle\mathcal{P}_{B2}\rangle$ and $\langle\mathcal{B}\rangle$ is very high across all scales. 

\textcolor{black}{The integral formula defining the non-local contribution $\langle\mathcal{B}^{{\rm nl}}\rangle$ makes this term very difficult to analyze and understand, including why this term also changes sign as $\ell$ is reduced. 
Nevertheless, the integral depends on the relative alignments of the filtered density- and velocity-gradients with respect to the vertical direction, much like the term $\langle\mathcal{P}_{B2}\rangle$. 
This suggests that the formation of ramp-cliffs might also be connected with the reversal of $\langle\mathcal{B}^{{\rm nl}}\rangle$, similar to the local flux. 
To further understand how the ramp-cliff structures correspond to the buoyancy flux reversals, we partition the flow based on the alignment of the filtered density gradient with respect to the vertical direction $\hat{\bm{e}}_{\tilde{B}} \cdot \ez$.
The more frequent ``ramps'' correspond to regions where the fluctuating density gradient misaligns with the mean density gradient $\hat{\bm{e}}_{\tilde{B}} \bm{\cdot} \ez >0$, whereas ``cliffs'' correspond to $\hat{\bm{e}}_{\tilde{B}} \cdot \ez <0$. 
We observe that cliff-events generate prominent forward buoyancy fluxes at all scales, converting TKE to TPE. 
At scales below the Ozmidov scale $\ell < l_O$, ramp-events produce strong reverse buoyancy flux events, converting TPE to TKE, whereas for $\ell \gg l_O$, ramp-events produce a forward flux. 
It is also quite remarkable that except for scales $\ell\leq O(\eta)$, the largest contributions to $\langle\mathcal{B}\rangle$ come from the extreme events where $\hat{\bm{e}}_{\tilde{B}} \bm{\cdot} \ez \approx -1$ and $\hat{\bm{e}}_{\tilde{B}} \cdot \ez \approx +1$, further emphasizing the crucial role that ramp-cliffs play in the buoyancy flux mechanism. 
The behaviour of the local and non-local part of the buoyancy flux conditioned on $\hat{\bm{e}}_{\tilde{B}} \bm{\cdot} \ez$ is qualitatively very similar, and only their relative contribution to the total buoyancy flux changes with filter scale. 
This provides significant evidence that the ramp-cliff mechanism that is responsible for the behaviour of the local buoyancy flux is also fundamentally that which is responsible for the behaviour of $\langle\mathcal{B}^{{\rm nl}}\rangle$ and its reversal in sign as $\ell$ is decreased.}

\textcolor{black}{These results} provide new insights for closure modelling of $\mathcal{B}$ in large eddy simulations that should be explored in future work, as well as trying to understand better the properties of $\mathcal{B}^{{\rm nl}}$.

\backsection[Acknowledgements]
{This research used resources of the Duke Compute Cluster at Duke University and the Oak Ridge Leadership Computing Facility at the Oak Ridge National Laboratory, which is supported by the Office of Science of the U.S. Department of Energy under Contract No. DE-AC05-00OR22725.  Additional resources were provided
through the U.S.\ Department of Defense High Performance Computing Modernization
Program by the Army Engineer Research and Development Center and the Army
Research Laboratory under Frontier Project FP-CFD-FY14-007.}

\backsection[Funding]
{S.B. and A.D.B. were supported by National Science Foundation (NSF) CAREER award \# 2042346. }

\backsection[Declaration of interests]
{The authors report no conflict of interest.}

\backsection[Author ORCIDs]{Soumak Bhattacharjee, https://orcid.org/0000-0002-3123-8973; Stephen M. de Bruyn Kops https://orcid.org/0000-0002-7727-8786; Andrew D. Bragg, https://orcid.org/0000-0001-7068-8048}

\bibliographystyle{jfm}
\bibliography{jfm}


\end{document}